\documentclass[prx,aps,showpacs,preprintnumbers,
amsmath,amssymb,superscriptaddress,longbibliography,colorlinks=true, a4paper=true, pdfstartview=FitV,linkcolor=blue, citecolor=blue, urlcolor=blue,twocolumn]{revtex4-1}

\usepackage[T1]{fontenc}		
\usepackage[english]{babel}		
\usepackage[utf8]{inputenc}	
\usepackage{times}
\usepackage{tikz}
\usepackage{tensor}
\usetikzlibrary{knots}

\usepackage{amsmath}			
\usepackage{amssymb}			
\usepackage{bbm}				

\usepackage{hyperref}
\usepackage{pbox}

\usepackage{graphicx}			
\usepackage{graphics}			
\DeclareGraphicsExtensions{.pdf}
\usepackage{color}	
\usepackage{makecell}

\newcommand{\bea}{\begin{eqnarray}}
\newcommand{\eea}{\end{eqnarray}}

\newcommand{\bk}{\mathbf{k}}

\newcommand{\bsigma} {\boldsymbol{\sigma}}

\DeclareMathOperator{\R}{\mathbb{R}}

\DeclareMathOperator{\C}{\mathbb{C}}

\DeclareMathOperator{\HH}{\mathcal{H}}

\newcommand{\bd}{\mathbf{d}}

\newcommand{\PT}{\mathcal{PT}}

\allowdisplaybreaks

\begin{document}
\title{
Fractal Nodal Band Structures}

\author{Marcus St{\aa}lhammar}
\affiliation{Nordita, KTH Royal Institute of Technology and Stockholm University, Hannes Alfv\'ens v\"ag 12, SE-106 91 Stockholm, Sweden}
\author{Cristiane Morais Smith}
\affiliation{Institute for Theoretical Physics, Utrecht University,
Princetonplein 5, 3584CC Utrecht, The Netherlands}

\begin{abstract}
Non-Hermitian systems exhibit interesting band structures, where novel topological phenomena arise from the existence of exceptional points at which eigenvalues and eigenvectors coalesce. One important open question is how this would manifest at noninteger dimension. Here, we report on the appearance of fractal eigenvalue degeneracies and Fermi surfaces in Hermitian and non-Hermitian topological band structures. This might have profound implications on the physics of black holes and Fermi surface instability driven phenomena, such as superconductivity and charge density waves. 
\end{abstract}
\maketitle


\section{Introduction}
The properties of fractal lines and surfaces comprise an important field within complex dynamics \cite{Gowrisankar2021} and topology \cite{Montiel1996,Porchon2021}, with wide applications in natural sciences. Prominent and popular examples include the shapes of Romanesco broccoli and seahorse tails, but it has also entered the stage in high-energy physics and the studies of black hole horizons \cite{nashie06}. Another promising venue is condensed-matter physics \cite{Ivaki2012}, where topology mainly has provided the classification of Bloch bands \cite{jantopreview} and the advent of topological insulators \cite{hasankane,qizhang} and semimetals \cite{goerbig,weylreview}. Recent years have, however, marked a paradigm shift within topological band theory as non-Hermitian (NH) systems have gained vastly increased attention \cite{NHreview}. Such models effectively describe dissipation or gain and loss, and find applications ranging from ultracold atoms \cite{Kreibich2014} and mechanical systems \cite{Ghatak2020} to optics \cite{Ozdemir2019,topphot,speclat,Ozawa2019}. 

A unique and ubiquitous feature of NH band structures is the generic appearance of exceptional points (EPs), where both eigenvalues and eigenvectors coalesce, leaving the corresponding Hamiltonian defective and thus, nondiagonalizable \cite{brody14}. The topological properties of EPs have been intensively studied, both in generic systems \cite{ourknots,Zhang2021,ourknots2}, but also in systems subject to discrete symmetries, where higher-order EPs are stabilized \cite{symprotnod,3EPemil,4EPMarcus,Sayyad22,Delplace21,Konig22}. Furthermore, novel topological phenomena has arisen following the studies of EPs \cite{Dembowski2001,Liu2021,Zhang2021}, and can be applied in sensing \cite{Hodaei2017,Chen2017,Lai2019,Chu2020,Wiersig2020,Yu2020,Peters2022} and unidirectional lasing \cite{Peng2014,Feng2014,Hodaei2014,Wang2021}, to cite just a few examples.

In this work, we merge studies on fractals and EPs by introducing a phase of matter: fractal nodal band structures. We show, through explicit constructions, that EPs can form Multibrot set boundaries in a range of different setups. Specifically, EPs of order two (EP2s) form fractal contours in generic three-dimensional (3D) systems, while they form contours and surfaces in NH parity-time ($\PT$)-symmetric systems in 2D and 3D, respectively. Quite remarkably, the corresponding Fermi surfaces (FSs) also take fractal shapes, as they terminate at the EPs. Additionally, we show that $\PT$ symmetry furthermore stabilize fractal contours of EP3s in NH systems, and contours of ordinary nodal points (ONPs) in Hermitian setups. 

The implications of our work are multifold. First, by providing concrete ways to construct fractal nodal FSs, we open the path to the realization of fractal superconductivity (SC) \cite{Fiegelman2010,Kravstov2012} and fractal charged density waves (CDWs), e.g., since these are FS instability phenomena, and hence, strongly depend on the shape of the FS. Second, it raises questions about black holes and analog gravity, as the fractal FSs in this work can be thought of as a tachyonic phase resembling the interior of a black hole with fractal horizon. Third, the number of concrete models hosting fractal nodal structures further suggests that our findings are of importance within various fields of experimental relevance, in addition to providing a missing piece in the understanding of both Hermitian and NH band structures in topological band theory.


\section{Exceptional fractals in 2D band structures}
The Multibrot set is defined as the Julia set given by the recursive relation $z_{n+1} = z_n^d+c$ for some complex $z_n$ and $c$. The corresponding boundary can be visualized using various methods including Jungreis functions and Newton methods \cite{Mandelbrot1982}. For the current purpose, the fractals are illustrated using Bernoulli lemniscates, or equipotential curves \cite{fractalbook}, where the boundary is depicted by plotting the absolute value of the function
\begin{equation} \label{eq:iteration}
    F(x,y,d) = f(x,y)+\left(f(x,y)+\left(f(x,y)+...\right)^d\right)^d = r
\end{equation}
with $f(x,y)=x+iy$, $r\in \C$ and for the Multibrot set, which will exemplify our findings, $d\geq 2$ and integer valued, with $d=2$ corresponding exactly to the Mandelbrot set. We note that our construction goes beyond $d\geq 2$, as $d$ in principle can be generalized to both negative and fractional values. For integer values, the generated images acquire a $|d-1|$-fold rotational symmetry. The fractal boundary of the Multibrot set can be thought of as a line embedding in a 2D space specifically given by
\begin{equation} \label{eq:fractalboundary}
\text{Re}\left[F(x,y,d)\right]^2+\text{Im}\left[F(x,y,d)\right]^2 = |r|^2.
\end{equation}

\begin{figure*}[t!]
\centering

\includegraphics[width=\textwidth]{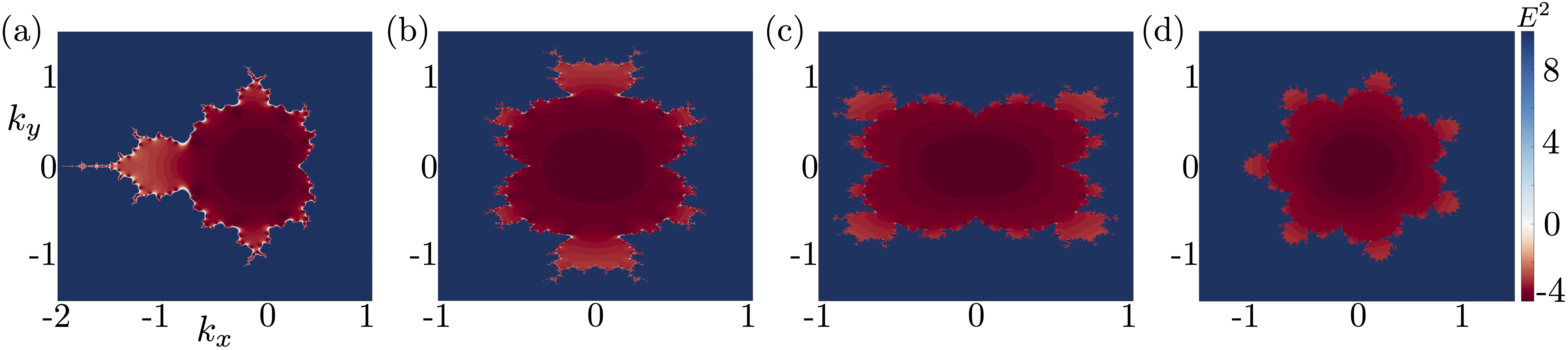}

\caption{\label{fig:PTcont} Fermi surfaces [the red region, defined as $\text{Re}\left(E\right)=0 \iff E^2<0$] and their bounding EPs (in white) of the $\PT$-symmetric model given by Eq.~\eqref{eq:dryPT2d}, taking the form of fractals. Specifically, they form Multibrot sets with $d=2,3,5,8$ in (a), (b), (c), and (d), respectively. In all panels, $r=2$, and Eq.~\eqref{eq:iteration} is terminated after ten steps. We emphasize that due to the topological properties of the eigenvalues, the EPs only comprise the line-shaped transition between the red ($\PT$ symmetry-broken) and blue ($\PT$-symmetric) region, and its seemingly different appearance in the various panels above is a consequence of the increasing energy gradient with $d$, in combination to numerical limitations.}
\end{figure*}

In terms of NH band structures, this is reminiscent of what is known for certain symmetry-protected exceptional structures. One such symmetry is $\PT$ symmetry, which due to its high experimental relevance will be the focus of the main text. We stress that similar features can be obtained using pseudo-Hermiticity, particle-hole-symmetry, or chiral symmetry \cite{Delplace21,Sayyad22}, and provide explicit models for those cases in Appendix~\ref{app:othersymm}. A $\PT$-symmetric Hamiltonian has to commute with the $\PT$ operator and consequently satisfies
$U_{\PT}\mathcal{H} U_{\PT}^{-1} = \mathcal{H}^*$, where $^*$ denotes complex conjugation and $U_{\PT}$ is a unitary matrix. Choosing $U_{\PT}=\mathbf{1}$, the matrix representation of a $\PT$-symmetric Hamiltonian consists of purely real entries. This will be the representation of $\PT$ symmetry used throughout this work. A $\PT$-symmetric two-band  Hamiltonian thus casts the form
\begin{equation} \label{eq:2bandPT}
    \HH_{\PT}^{2b} = d_{R,x}\sigma^x + i d_{I,y} \sigma^y + d_{R,z}\sigma^z,
\end{equation}
where $\sigma^i$, $i=x,y,z$ denotes Pauli matrices, and $d_{R,x},d_{I,y}$ and $d_{R,z}$ are real-valued continuously differentiable functions of the lattice momentum components $k_x,k_y,$ and $k_z$. As the current study focuses on eigenvalue degeneracies, the term proportional to the identity matrix has been neglected in Eq.~\eqref{eq:2bandPT}, as such a term will merely shift the EPs in energy space. The corresponding eigenvalues are $E_{\PT}^{2b,\pm} = \pm \sqrt{d_{R,x}^2+d_{R,z}^2-d_{I,y}^2}$, and EP2s appear when
\begin{equation} \label{eq:PTEP2}
    d_{R,x}^2+d_{R,z}^2=d_{I,y}^2.
\end{equation}
The similarities between Eqs.~\eqref{eq:PTEP2} and \eqref{eq:fractalboundary}, indicate that the EP2s ought to form the boundary of a Multibrot set if the corresponding Hamiltonian is defined as
\begin{equation}
    d_{R,x} = \text{Re}\left(F\right), \quad d_{R,z} = \text{Im}\left(F\right), \quad d_{I,y} = r \label{eq:dryPT2d},
\end{equation}
where we will use $F:=F(k_x,k_y,d)$ throughout for simplicity. The resulting exceptional fractals and their corresponding FSs (which take the form of the Multibrot set itself) are displayed in Fig.~\ref{fig:PTcont} for different values of $d$. Here we follow the convention of Ref.~\cite{ourknots}, and define the FS as $\text{Re}\left(E\right)=0$. Apart from these polynomial models, fractallike exceptional contours also appear in periodic models, as shown in Appendix~\ref{app:permod}.


\section{3D extensions of fractal contours of EPs}
When increasing the spatial dimension from two to three, different things will happen depending on whether $\PT$ symmetry is preserved or not. We here consider several cases, all implying straightforward generalizations of the fractal boundaries into embeddings in 3D.

\begin{figure}[b!]
\centering

\includegraphics[width=\columnwidth]{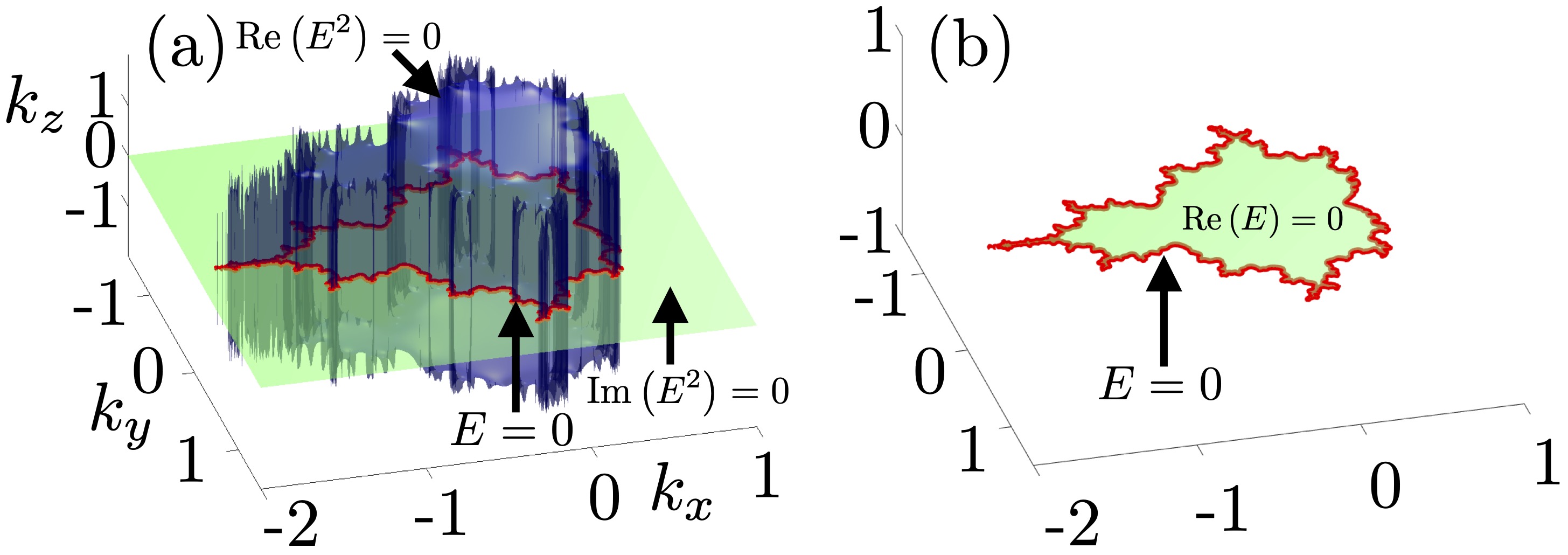}

\caption{\label{fig:3DmodelsGeneric} Fractal nodal band structures for the generic NH two-band model given by Eq.~\eqref{eq:im2bandgen}. Panel (a) illustrates how the EP2s (red curve) appear as the intersection between two surfaces, defined by  $\text{Re}\left(E^2\right)=0$ (blue) and $\text{Im}\left(E^2\right)=0$ (green). Panel (b) shows the corresponding FS. This provides a straightforward 3D generalization of the notion of the boundary of the Mandelbrot set, i.e., as the embedding of a contour in 3D space. Here, $d=2$, $r=2$, and Eq.~\eqref{eq:iteration} is terminated after ten steps.}
\end{figure}

In the absence of any symmetries, EP2s are of codimension two and will therefore form contours in 3D. A generic such two-band model can be written as $\HH^{2b} = \bd_R\cdot \bsigma+i \bd_I\cdot \bsigma$, where $\bd_R,\bd_I:\R^3\to \R^3$ are continuously differentiable functions of lattice momentum and $\bsigma$ denotes the vector of Pauli matrices. The corresponding EPs occur as the intersection between two implicitly defined surfaces given by  
\begin{equation}
    \text{Re}\left(E^2\right) = \bd_R^2-\bd_I^2=0, \quad \text{Im}\left(E^2\right) = 2\bd_R\cdot \bd_I=0.
\end{equation}
The solutions to these equations will form fractals when they are of the form given in Eq.~\eqref{eq:fractalboundary}. One such example is provided by
\begin{equation} 
    \bd_R=\left[\text{Re}\left(F\right),\text{Im}\left(F\right),g(k_z)\right], \quad \bd_{I}=\left(0,0,r\right),\label{eq:im2bandgen}
\end{equation} 
where $F$ is defined in Eq.~\eqref{eq:iteration}, and $g(k_z)$ is a continuously differentiable function depending only on $k_z$. The EPs then corresponds to regions where $g(k_z)=0$ and $\text{Re}\left(F\right)^2+\text{Im}\left(F\right)^2=r^2$, simultaneously. Figures.~\ref{fig:3DmodelsGeneric}(a) and \ref{fig:3DmodelsGeneric}(b) exemplify one such solution, indicating that they form fractal contours embedded in 3D. As before, the FS,  depicted in Fig.~\ref{fig:3DmodelsGeneric}(b), forms the corresponding Multibrot set.


\begin{figure}[hbt!]
\centering

\includegraphics[width=\columnwidth]{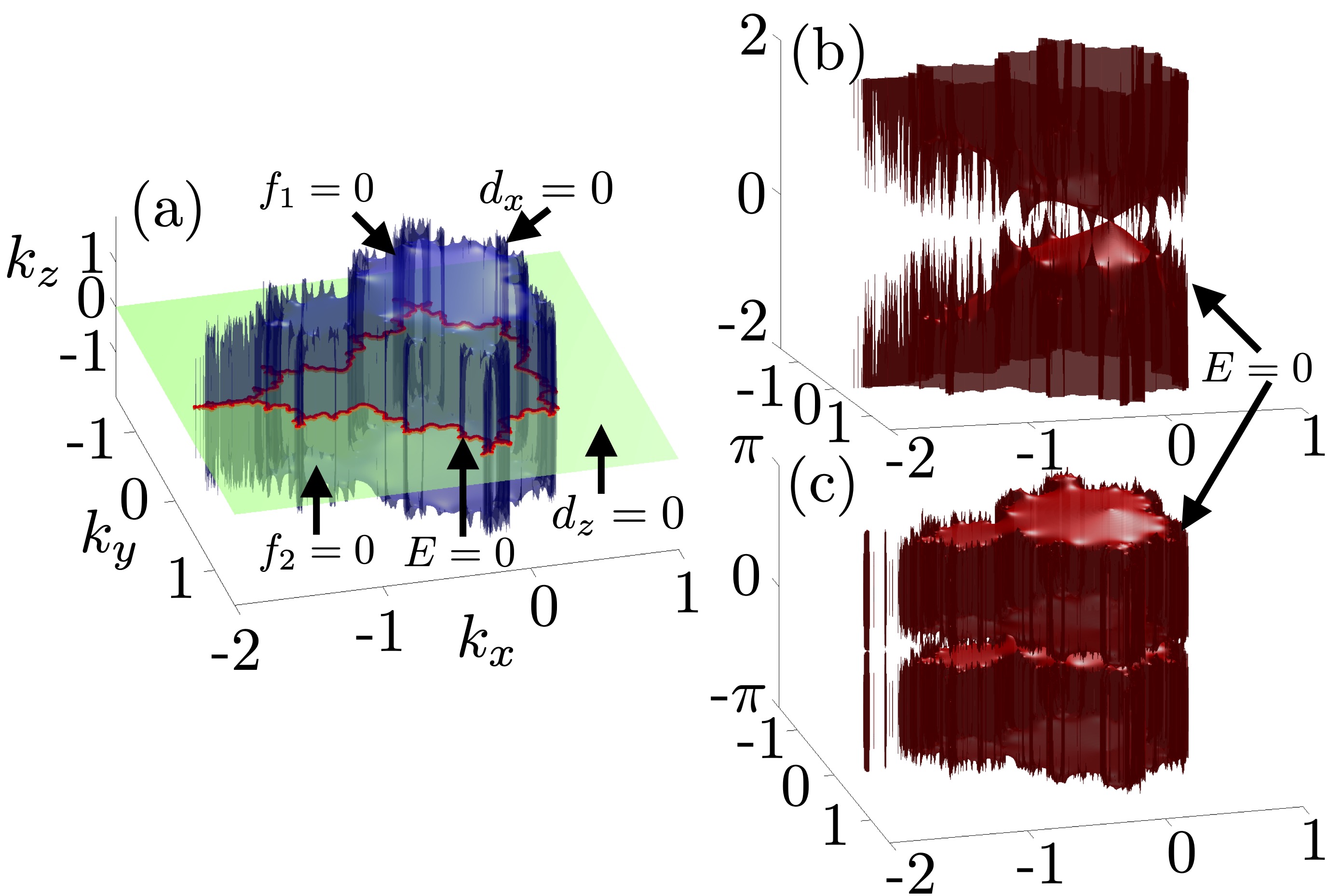}

\caption{\label{fig:3DmodelsPT} Fractal nodal band structures for various $\PT$-symmetric systems. Panel (a) displays the EP3s and ONPs (red curve) of the systems given by Eq.~\eqref{eq:PT3band} and Eq.~\eqref{eq:Herm2band}, respectively. Consequently, the blue surface illustrates either solutions to $f_1=0$ or $d_x=0$, while the green surface corresponds to either $f_2=0$ or $d_z=0$. Panels (b) and (c) illustrate the EP2s of a 3D $\PT$-symmetric Hamiltonian defined by Eq.~\eqref{eq:dryPT2d}, with $r$ interchanged for $k_z$ (b) and $\sin k_z$ (c), respectively. These fractallike surfaces intersect each other at points known as nondefective EPs, in close analogy to how energy bands in Weyl semimetals host point-like intersections. Here, $d=2$ in all panels, $r=2$ in panel (a), and Eq.~\eqref{eq:iteration} is terminated after ten steps.}
\end{figure}

\section{Exceptional fractals in 3D $\PT$-symmetric systems}
If the third dimension is instead introduced in a way that preserves $\PT$ symmetry, the contours of EP2s are promoted to surfaces accompanied by contours of EP3s and pointlike EP4s. Thus, it is possible that fractal contours appear as threefold eigenvalue degeneracies. Inspired by the construction introduced in Ref.~\cite{4EPMarcus}, we consider a Hamiltonian represented by
\begin{equation} \label{eq:PT3band}
    \HH_{\PT}^{3b} = \begin{pmatrix}
    f_2&0&\Lambda
    \\ 
    \alpha & 0 & \beta 
    \\ 
    f_2 & f_1 & -f_2    \end{pmatrix},
\end{equation}
with $\alpha,\beta,\Lambda \in \R$. The eigenvalues (here denoted $\lambda$) can then be determined by solving the characteristic equation
\begin{equation}
    f_1\left(\alpha \Lambda -\beta f_2\right)+\lambda\left(f_2^2+\Lambda f_2 + \beta f_1\right)-\lambda^3 = 0.
\end{equation}
Threefold eigenvalue degeneracies correspond to regions in parameter space where both the constant term and the coefficient of the linear term in $\lambda$ vanish. For $|\Lambda|$ sufficiently large, the solutions are given by $f_1=f_2=0$, and by choosing, e.g., $f_1=\text{Re}\left(F\right)^2+\text{Im}\left(F\right)^2-r^2$ and $f_2=g(k_z)$, the corresponding EP3s form fractal contours. One such example is displayed in Fig.~\ref{fig:3DmodelsPT}(a), which is of the exact same form as the structure displayed in Fig.~\ref{fig:3DmodelsGeneric}(a), where the red contour is a collection of EP2s rather than EP3s, as is the case of Fig.~\ref{fig:3DmodelsPT}(a).

Apart from appearing as exceptional eigenvalue degeneracies in NH systems, nodal fractals can also appear in Hermitian band structures. Building further on the various constructions presented above, a concrete example hosting nodal fractals is given by the following 3D $\PT$-symmetric Hermitian two-band model,
\begin{equation} \label{eq:Herm2band}
    \HH^{2b}_{\PT,H} = d_x\sigma^x + d_z\sigma^z,
\end{equation}
with $d_x =\text{Re}\left(F\right)^2+\text{Im}\left(F\right)^2-r^2$ and $d_z = g(k_z)$. The eigenvalue degeneracies occur when $d_x$ and $d_z$ vanish simultaneously, and will thus correspond to fractal contours embedded in 3D momentum space. These are shown in Fig.~\ref{fig:3DmodelsPT}(a). We again note the similarities between Fig.~\ref{fig:3DmodelsGeneric}(a) and Fig.~\ref{fig:3DmodelsPT}(a), and emphasize the physical difference in interpreting the pictures, since the red contour in Fig.~\ref{fig:3DmodelsPT}(a) is now to be interpreted as ONPs, rather than EPs.

Lastly, it should furthermore be possible to obtain a fractal-like surface of EP2s. A naive attempt would be to use a Hamiltonian very similar to the one defined in Eq.~\eqref{eq:dryPT2d}, but exchanging $r$ for $k_z$. This will result in a model whose EP2s form ``fractals'', in the sense that a given slice in $k_z$ host EP2s similar to those displayed in Fig.~\ref{fig:PTcont}. This surface however extends to infinity, which is depicted in Fig.~\ref{fig:3DmodelsPT}(b), since $k_z$ is an unbounded function. Alternatively, the third dimension can be introduced periodically, which at least naively, would correspond to couple sheets of 2D models together in a NH way. One such option is displayed in Fig.~\ref{fig:3DmodelsPT}(c), where $k_z$ is exchanged for $\sin k_z$. This results in two finite fractal-like surfaces of EP2s. It should furthermore be noted that both the surfaces displayed in Figs.~\ref{fig:3DmodelsPT}(b) and \ref{fig:3DmodelsPT}(c) intersect each other at the origin [and at $\bk = (0,0,\pi)$]---an intersection that is stable in the sense that it is protected by the symmetry. Such intersection points have been studied earlier and are sometimes referred to as nondefective EPs \cite{Sayyad22a}.

\begin{figure}[t!]
\centering

\includegraphics[width=\columnwidth]{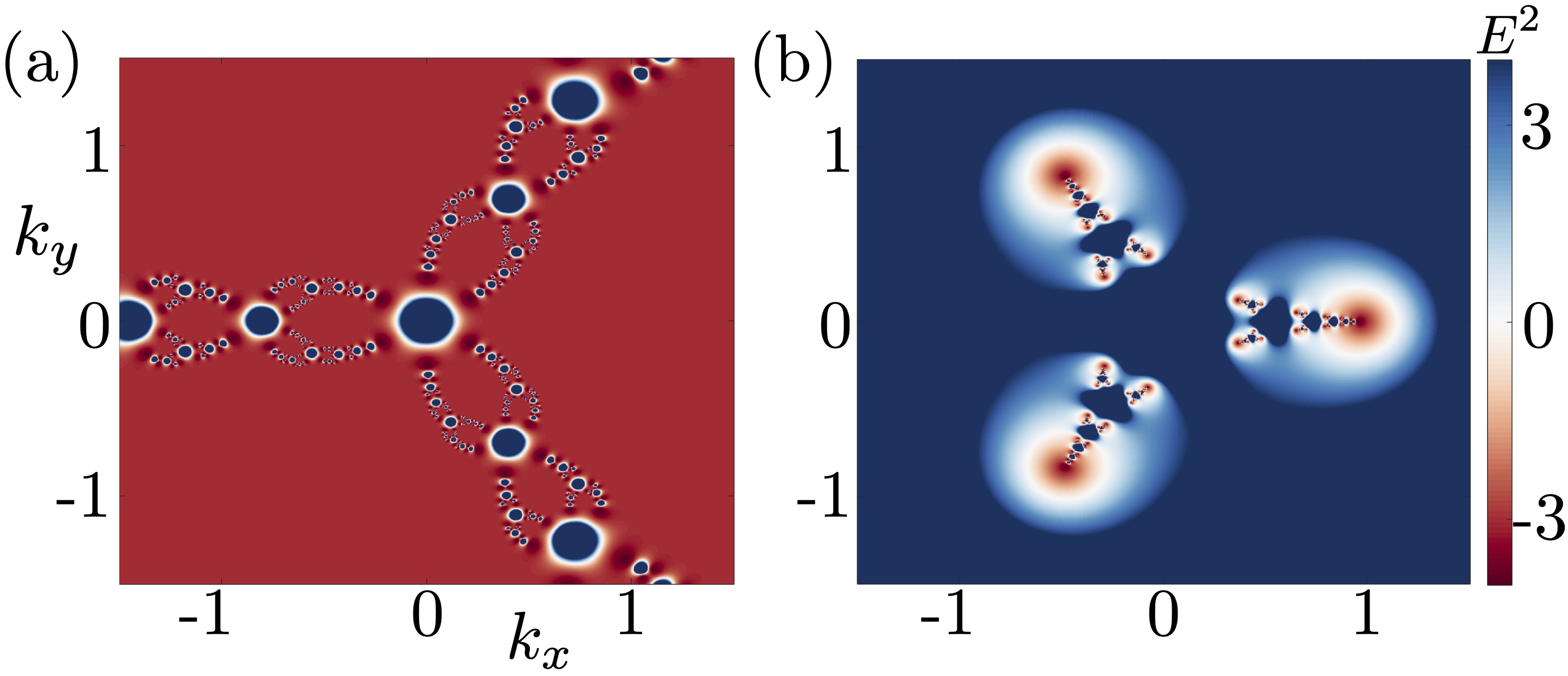}

\caption{\label{fig:Newton} Exceptional Newton fractals in 2D $\PT$-symmetric models, given by Eq.~\eqref{eq:2bandPT}, obtained by substituting Eq.~\eqref{eq:Newton} into Eq.~\eqref{eq:fractalboundary} using five iterations and $r=2$. In both panels, $p(z)=z^3-1$, while $a=1$ in panel (a) and $a=-\frac{1}{2}$ in panel (b). This illustrates that our construction goes beyond the Multibrot set, and can be applied for any algebraic fractal.}
\end{figure}

\begin{figure*}[htb!]
\centering

\includegraphics[width=\textwidth]{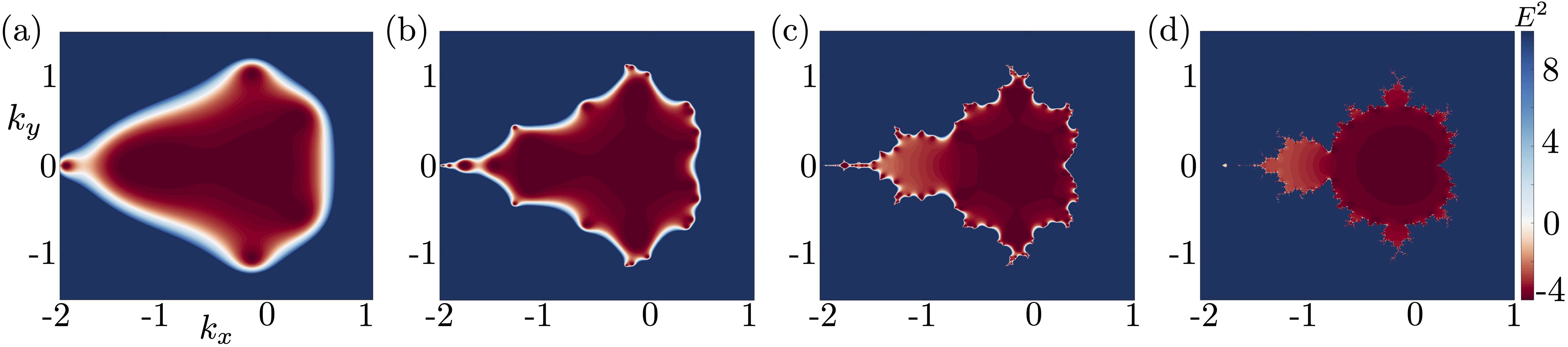}

\caption{\label{fig:Transitions} Fermi surfaces [corresponding to $E^2<0$] and their bounding EP2s in $\PT$-symmetric models given by Eq.~\eqref{eq:dryPT2d}, with $d=2$ and $r=2$. The different panels illustrate how the approximation in terms of lemniscates affect the form of the EP2s. The number of iterations of Eq.~\eqref{eq:iteration} is 3, 5, 8 and 18 in (a), (b), (c) and (d), respectively, corresponding to approximating the boundary of the Mandelbrot set with 3, 5, 8, and 18 lemniscates, respectively. As the complexity of the model increases with the number of iterations, it is more likely that EPs and FSs measured in experiments take forms similar to those displayed in (a) or (b), rather than (c) or (d). Thus, the observations in experiments are not expected to reproduce perfect fractals, but rather a contour reminiscent of a fractal approximation.}
\end{figure*}

\section{Discussion}
In this work, we extended the plethora of eigenvalue degeneracies in both Hermitian and NH systems to include fractal-like structures. Through specific constructions, we showed that boundaries of Multibrot sets appear as stable contours of EP2s in 2D systems subject to $\PT$ symmetry, in generic systems in 3D, and as stable ONPs in Hermitian systems subject to $\PT$ symmetry in 3D. Furthermore, we illustrated in specific examples how NH $\PT$-symmetric systems also generically host fractal contours of EP3s as well as fractallike surfaces of EP2s in 3D.

The nodal fractals and fractal FSs are as stable in the sense that they are protected by the underlying symmetry. This means that the fractal structures are stable toward any small and symmetry-preserving perturbation to the parent Hamiltonian, but they will dissolve if a symmetry-breaking perturbation is added. In Appendix~\ref{app:stab}, we provide details and figures further supporting this claim.

Our method is not limited to the Multibrot sets, but can in principle be utilized to achieve systems whose nodal structures correspond to any algebraic fractal, i.e., a fractal that can be described in terms of some function. To illustrate this generality, we present a few examples of exceptional Newton fractals in 2D $\PT$-symmetric systems. Newton fractals are defined as the Julia set of
\begin{equation} \label{eq:Newton}
z_{n+1}= z_n-a \frac{p(z_n)}{p'(z_n)},
\end{equation}
where $a, p(z)\in \C$, and $p'(z)$ is the derivative of $p(z)$. The Newton fractals shown in Fig.~\ref{fig:Newton} merely comprise an additional example of structures arising from this method, but yet again illustrate the potential of the presented arguments.

The large variety of models hosting nodal fractals suggests an equally wide range of experimental setups for their potential realization. The presence of $\PT$ symmetry implies that photonic crystals are good candidates for the observation of fractal nodal structures both in 2D and 3D, where EPs can be directly imaged using photoluminescence measurements \cite{kim2016} or angle-resolved thermal emission spectroscopy \cite{NHPC}. For Hermitian systems, nodal lines have been observed in the semimetal ZrB$_2$ \cite{Lou2018} and Mg$_3$Bi$_2$ \cite{Chang2018}, along with nodal links in TiB$_2$ \cite{Liu2018}. Hence, similar materials are natural experimental candidates for more complicated nodal line structures. Recent studies further suggest that phononic crystals are also of significance, as they  can host nodal rings \cite{Deng2019}, links \cite{Wang2022}, and even chains \cite{Lu2020}. The  complexity of systems hosting nodal fractals will, however, require high-precision tuning of the studied system, a hurdle that could be difficult to overcome in the above mentioned examples. Thus, a more promising candidate is  single-photon interferometry, a technique that provides the ability to tune system parameters with remarkable precision and has been utilized to realize knotted EP2s in 3D and $\PT$ symmetry-protected second-order exceptional rings in 2D \cite{expknots}. This indicates that it could be used to realize the fractal EPs and FSs displayed in Figs.~\ref{fig:PTcont}, \ref{fig:3DmodelsGeneric} and \ref{fig:Newton}. Very recently, using the same technique, the authors of Ref.~\cite{Wang2023} were further able to realize EP3s in $\PT$ and chirally symmetric systems in 2D, suggesting that also fractal contours of EP3s and surfaces of EP2s, such as those displayed in Fig.~\ref{fig:3DmodelsPT}, could be probed by modifying the current setup. It should be noted that even though it will be impossible to observe a perfect fractal (as it would require an approximation including infinitely many lemniscates), fractallike nodal eigenvalue degeneracies are experimentally feasible. In Fig.~\ref{fig:Transitions}, we illustrate the transition into a more fractallike exceptional contour in the 2D $\PT$-symmetric systems given by Eq.~\eqref{eq:dryPT2d}, indicating what one would expect to see in experiments.

A more phenomenological, yet interesting, observation is a possible connection between fractal nodal band structures and analog gravity, where recent studies have unraveled a relation between the structures formed by the EPs in $\PT$-symmetric systems and certain black hole setups \cite{stalhammar21}. Although the main focus there has been on observable analogs of Hawking radiation, a link to gravitational systems is interesting since the NH $PT$-symmetric FSs in 2D can be thought of as a tachyonic phase, as it corresponds to a region where the square energy is negative in $\PT$-symmetric setups \cite{Feinberg1967}. Similar case-studies have been carried out in other aspects than analog gravity, unraveling the transport properties of such tachyonic phases in NH setups \cite{Sticlet2021}. Taking into account that certain black hole studies indicate that event horizons host fractal structures \cite{nashie06}, it is intriguing to think of the possibility for the exceptional fractals to represent an analog black hole horizon, with the corresponding interior mimicked by the tachyonic phase comprised of the FS.

The unraveling of fractal phases in band theory opens up several further questions and research directions. Concrete examples include the understanding of what it means for a system to host FSs with infinite perimeter, and what physical role is played by the corresponding Hausdorff (or fractal) dimension. The latter is interesting from another point of view, as it is related to how these fractal structures are characterized. This is not obvious as one could argue that they are all topologically equivalent to circles or spheres. This suggests that a classification scheme going beyond topological classification is necessary to fully understand fractal nodal band structures.

Another very intriguing open question is the fate of FS instability phenomena, such as SC and CDW in fractal FS. The self-similarity of fractals implies that Cooper pairing, e.g., might occur recursively in self-similar parts of the FS, leading to a cascade of momenta pairing. Should one then observe fractal SC \cite{Fiegelman2010,Kravstov2012}? Which new features could the self-similarity of the momenta bring? Furthermore, competing phenomena, such as CDW, which is rooted on the nesting $\bk$-vector arising repeatedly at the FS could give rise to beatings of the self-similar length scales, in addition to rotation. More interestingly, there should be a diverging density of states at the boundaries, since the perimeter of a fractal tends to infinity. This infinite perimeter may also lead to new interesting phenomena related to eigenstates, in particular, their collective coalescence.
Finally, on a more generic level, this work raises the question whether one could enlarge the portfolio of qubits and qutrits into qufractits, and opens a path for a large variety of phenomena unforeseen to present.

\section*{Acknowledgements}
The authors acknowledge fruitful discussions with L. R\o dland and V. Gritsev. CMS acknowledges the research program “Materials for the Quantum Age” (QuMat) for financial support. This program (Registration No. 024.005.006) is part of the Gravitation Program financed by the Dutch Ministry of Education, Culture and Science (OCW).

\newpage

\appendix


\section{Symmetry-protected fractal contours of EPs}

\subsection{Generalization to additional symmetries} \label{app:othersymm}
\begin{figure*}[t!]
\centering

\includegraphics[width=\textwidth]{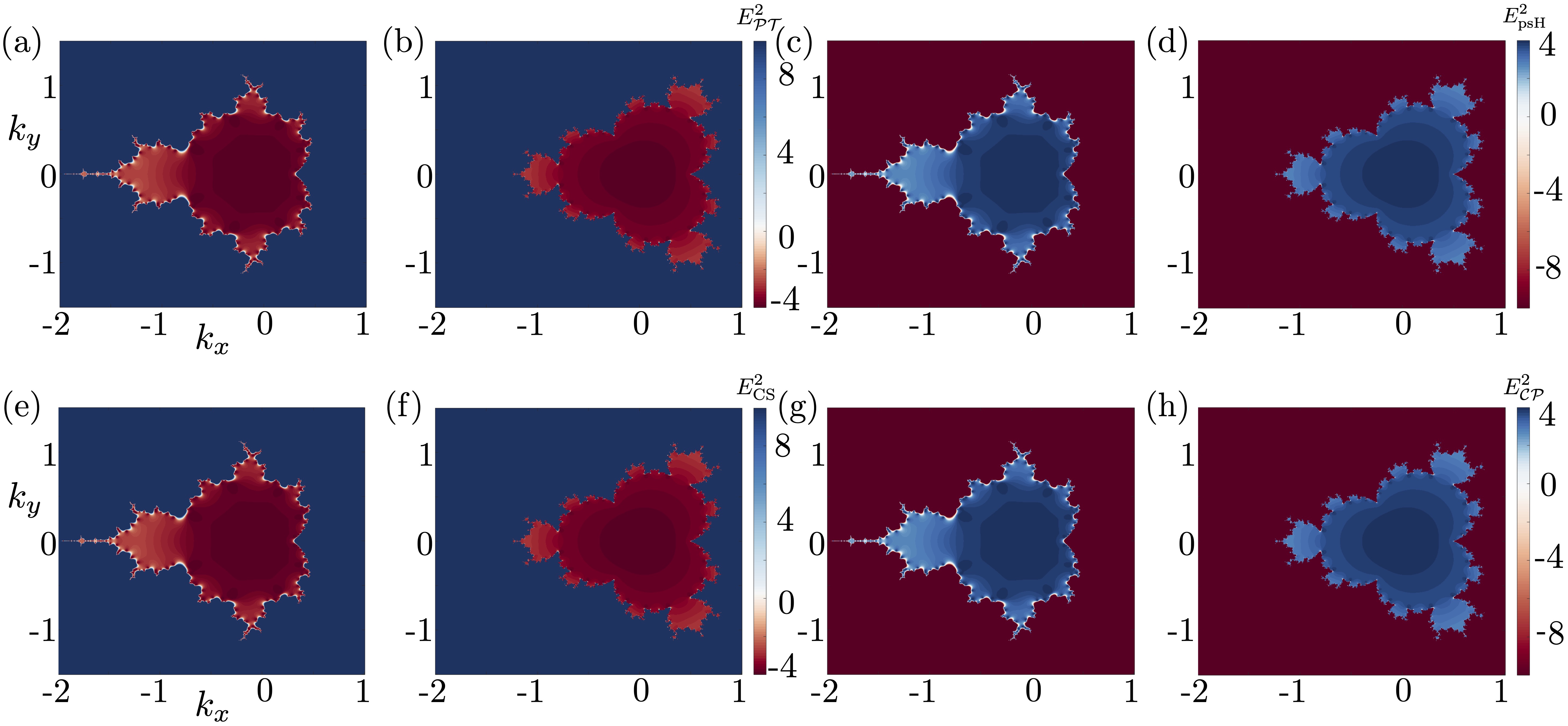}

\caption{\label{fig:Gensymm} Fermi surfaces [defined as $\text{Re}\left(E\right)=0 \iff E^2<0$] and their bounding EPs of the models given by Eq.~\eqref{eq:PTHamsym} in panels (a), (b); Eq.~\eqref{eq:PSHHamsym} in panels (c), (d); Eq.~\eqref{eq:CSHamsym} in panels (e), (f); and Eq.~\eqref{eq:CPHamsym} in panels (g), (h), taking fractallike forms. In panels (a), (c), (e), and (g)  $d=2$, and $d=3$ in the remaining panels. The different symmetries give rise to very similar FS structures, with the important difference being that the meaning of the FS and imaginary FS [defined as $\text{Im}\left(E\right)=0\iff E^2>0$] changes in psH and $\mathcal{CP}$ systems, as compared to $\PT$ and CS systems. In all panels $r=2$, and Eq.~\eqref{eq:iteration} is terminated after ten steps.}
\end{figure*}

A requirement for a symmetry to possibly allow for the existence of stable fractal contours of EPs in 2D, is that it reduces the number of constraints defining the location of the EPs. Following the works in Refs.~\cite{Delplace21,Sayyad22}, e.g., the relevant symmetries are $\PT$ symmetry (as studied in the main text), pseudo-Hermiticity (psH), chiral symmetry (CS) and  particle-hole symmetry ($\mathcal{CP}$). Operators subject to these symmetries are bound to satisfy
\begin{align}
    U_{\PT}\HH_{\PT}U^{-1}_{\PT} &=H^*_{\PT}, \label{eq:PTconstraint}
    \\
    U_{\text{psH}}\HH_{\text{psH}}U^{-1}_{\text{psH}} &= H^{\dagger}_{\text{psH}},
    \\
    U_{\text{CS}}\HH_{\text{CS}}U^{-1}_{\text{CS}} &= -\HH^{\dagger}_{\text{CS}},
    \\
    U_{\mathcal{CP}}\HH_{\mathcal{CP}}U^{-1}_{\mathcal{CP}} &= -H^*_{\mathcal{CP}},
\end{align}
respectively. Here, $U_A$, $A\in\left\{\PT,\text{psH},\text{CS},\mathcal{CP}\right\}$ denote unitary operators. To illustrate the appearance of fractal contours of EPs in these systems, we now consider specific two-band systems. Using the representation $U_{\PT} = \sigma^0$, $U_{\text{psH}}=U_{\mathcal{CP}}=\sigma^x$, and $U_{\text{CS}}=\sigma^z$, the respective twp-band Hamiltonians will cast the form
\begin{align}
    \HH_{\PT}&=d^{\PT}_{R,x}\sigma^x+id^{\PT}_{I,y}\sigma^y+d^{\PT}_{R,z}\sigma^z,
    \\
    \HH_{\text{psH}} &= d^{\text{psH}}_{R,x}\sigma^x+id^{\text{psH}}_{I,y}\sigma^y+id^{\text{psH}}_{I,z}\sigma^z,
    \\
    \HH_{\text{CS}} &=d^{\text{CS}}_{R,x}\sigma^x+d^{\text{CS}}_{R,y}\sigma^y+id^{\text{CS}}_{I,z}\sigma^z,
    \\
    \HH_{\mathcal{CP}} &= id^{\mathcal{CP}}_{I,x}\sigma^x+id^{\mathcal{CP}}_{I,y}\sigma^y+d^{\mathcal{CP}}_{R,z}\sigma^z.
\end{align}
Thus, the corresponding EP2s will be given as solutions to the following equations:
\begin{align}
    E^2_{\PT}=0&\iff\left(d^{\PT}_{R,x}\right)^2+\left(d^{\PT}_{R,z}\right)^2=\left(d^{\PT}_{I,y}\right)^2,
    \\
    E^2_{\text{psH}}=0&\iff\left(d^{\text{psH}}_{I,y}\right)^2 + \left(d^{\text{psH}}_{I,z}\right)^2 =\left(d^{\text{psH}}_{R,x}\right)^2,
    \\
    E^2_{\text{CS}}=0&\iff\left(d^{\text{CS}}_{R,x}\right)^2 + \left(d^{\text{CS}}_{R,y}\right)^2 =\left(d^{\text{CS}}_{I,z}\right)^2,
    \\
    E^2_{\mathcal{CP}}=0&\iff\left(d^{\mathcal{CP}}_{I,x}\right)^2+\left(d^{\mathcal{CP}}_{I,y}\right)^2=\left(d^{\mathcal{CP}}_{R,z}\right)^2.
\end{align}
These will take the form of fractals if the Hamiltonian is defined as, e.g.,
\begin{align}
    \HH_{\PT} &= \text{Re}\left[F\left(k_x,k_y,d\right)\right]\sigma^x+\text{Im}\left[F\left(k_x,k_y,d\right)\right]\sigma^z+ir\sigma^y \label{eq:PTHamsym},
    \\
    \HH_{\text{psH}} &= i\text{Re}\left[F\left(k_x,k_y,d\right)\right]\sigma^y+i\text{Im}\left[F\left(k_x,k_y,d\right)\right]\sigma^z+r\sigma^x, \label{eq:PSHHamsym}
    \\
    \HH_{\text{CS}} &= \text{Re}\left[F\left(k_x,k_y,d\right)\right]\sigma^x+\text{Im}\left[F\left(k_x,k_y,d\right)\right]\sigma^y+ir\sigma^z, \label{eq:CSHamsym}
    \\
    \HH_{\mathcal{CP}} &= i\text{Re}\left[F\left(k_x,k_y,d\right)\right]\sigma^x+i\text{Im}\left[F\left(k_x,k_y,d\right)\right]\sigma^y+r\sigma^z. \label{eq:CPHamsym}
\end{align}
The resulting squared spectra are displayed in Fig.~\ref{fig:Gensymm} for various values of $d$ and $r$. The EPs take the same forms for all these different models, but notably, the meaning of the FS and the imaginary FS (which is defined as $\text{Im}\left(E\right)=0 \iff E^2>0$ in these systems, following Ref.~\cite{ourknots}) are exchanged in systems subject to psH or $\mathcal{CP}$ symmetry, compared to those subject to $\PT$ symmetry or CS. 

\subsection{Symmetry-protection and stability} \label{app:stab}
The symmetry-protected nodal fractals are highly dependent on the underlying symmetry of the system: if any arbitrarily small, but finite, symmetry-breaking perturbation is added to the parent Hamiltonian, the nodal fractal immediately disappears. To illustrate that the nodal fractals are indeed stable and preserved as long as the relevant symmetry is preserved, we will here investigate the impact of both symmetry-preserving and symmetry-breaking perturbations. As this study is completely analogous for all the symmetries listed above, as well as for the Hermitian version of $\PT$-symmetry, we will restrict the discussion to NH $\PT$-symmetric systems only, recalling that similar arguments can be applied for the other symmetries as well. It should also be noted that the reasoning regarding stability also holds for fractal contours of EPs in generic NH systems.

\begin{figure*}[hbt!]
\centering

\includegraphics[width=\textwidth]{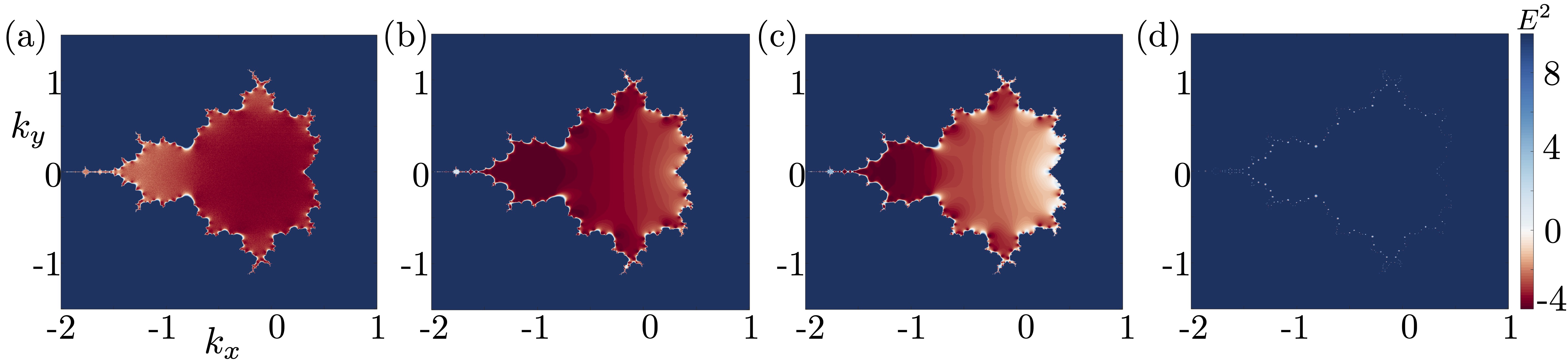}

\caption{\label{fig:Perturbation} Fermi surfaces [defined as $\text{Re}\left(E\right)=0 \iff E^2<0$] and their bounding EPs of the $\PT$-symmetric model given by Eq.~\eqref{eq:unpertham} with various perturbations added. (a) A perturbation of the form $10^{-1}\left(\alpha_i + \beta_ik_x+\gamma_ik_y\right)\sigma^i$, with $i=x,y,z$ summed over, and $\alpha_i, \beta_i$, and $\gamma_i$ randomly generated numbers between zero and one is added. Despite adding this perturbation, the fractal structure of the EPs and the FS is intact. (b)--(d) A perturbation of the form $\delta\sigma^x$ is added, where $\delta=1,1.5,5$ in (b), (c), and (d), respectively. Here, the fractal structure is gradually dissolved, and eventually results in a cluster of small circles of EPs bounding disk-shaped FSs. This shows that the fractal structure is stable under small enough symmetry-preserving perturbations, which means that it is protected by the underlying symmetry. In all panels, $r=2$, $d=2$, and Eq.~\eqref{eq:iteration} is terminated after ten steps.}
\end{figure*}

Consider the NH $\PT$-symmetric two-band model given by
\begin{equation}\label{eq:unpertham}
    \HH=\text{Re}\left[F\left(k_x,k_y,2\right)\right]\sigma^x+\text{Im}\left[F\left(k_x,k_y,2\right)\right]\sigma^z+ir\sigma^y,
\end{equation}
the squared spectrum of which is plotted in Fig.~\ref{fig:Gensymm}(a) for $r=2$. We now want to investigate what happens when we add small, symmetry-preserving perturbations to this system. For the sake of simplicity, we will allow one of the perturbations to have constant terms and terms linear in momentum components, while the other perturbation is kept constant. However, we stress that the findings are valid for more general perturbations as well. Fig.~\ref{fig:Perturbation}(a) shows that the spectrum still exhibits a fractal FS and that the EPs form a fractal contour in momentum space when applying a perturbation of the form $10^{-1}\left(\alpha_i + \beta_ik_x+\gamma_ik_y\right)\sigma^i$, where $i=x,y,z$ is summed over, and $\alpha_i, \beta_i$ and $\gamma_i$ are randomly generated numbers between zero and one. In Fig.~\ref{fig:Perturbation}(b)--(d), a constant perturbation $\delta\sigma^x$ is added to Eq.~\eqref{eq:unpertham}, showing how strong the perturbation must be to break the fractal structure of the EPs and the FS. In (b) $\delta=1$ and (c) $\delta=1.5$, and the fractality remains intact, while in (d), where $\delta=5$, it is ruined. The EPs now form a cascade of circles that bounds a family of disk-shaped FSs. This means that the fractal FSs and EPs are intact in a finite region of the symmetry-preserving perturbation-strength. Our findings therefore do not rely on fine-tuning of system parameters, but rather on the ability to implement relevant symmetries in NH setups.

It is worth recalling that an arbitrarily small but finite symmetry-breaking perturbation, i.e., a perturbation that does not satisfy Eq.~\eqref{eq:PTconstraint} in the case of $\PT$ symmetry, will immediately ruin the fractal structure, as the resulting EPs will be of codimension two instead of one, making them form points instead of contours in 2D momentum space. This can be visualized straight-forwardly by investigating how the equations defining the EPs change upon adding such a perturbation. Adding such a term, e.g., $\epsilon\sigma^y$, $\epsilon\in \R\setminus \left\{0\right\}$ to Eq.~\eqref{eq:unpertham} will give EPs corresponding to the solutions to,
\begin{align}
    \text{Re}\left[F(k_x,k_y,2)\right]^2+\text{Im}\left[F(k_x,k_y,2)\right]^2+\epsilon^2 &= r^2,
    \\
    2i\epsilon r &= 0. \label{eq:gapped2}
\end{align}
Since both $r$ and $\epsilon$ are taken to be real non-zero constants, Eq.~\eqref{eq:gapped2} is never fulfilled. Thus, all the EPs are gapped out, and the fractal structure is completely dissolved.

\section{Fractal eigenvalue degeneracies in periodic models} \label{app:permod}
\begin{figure}[hbt!]
\centering

\includegraphics[width=\columnwidth]{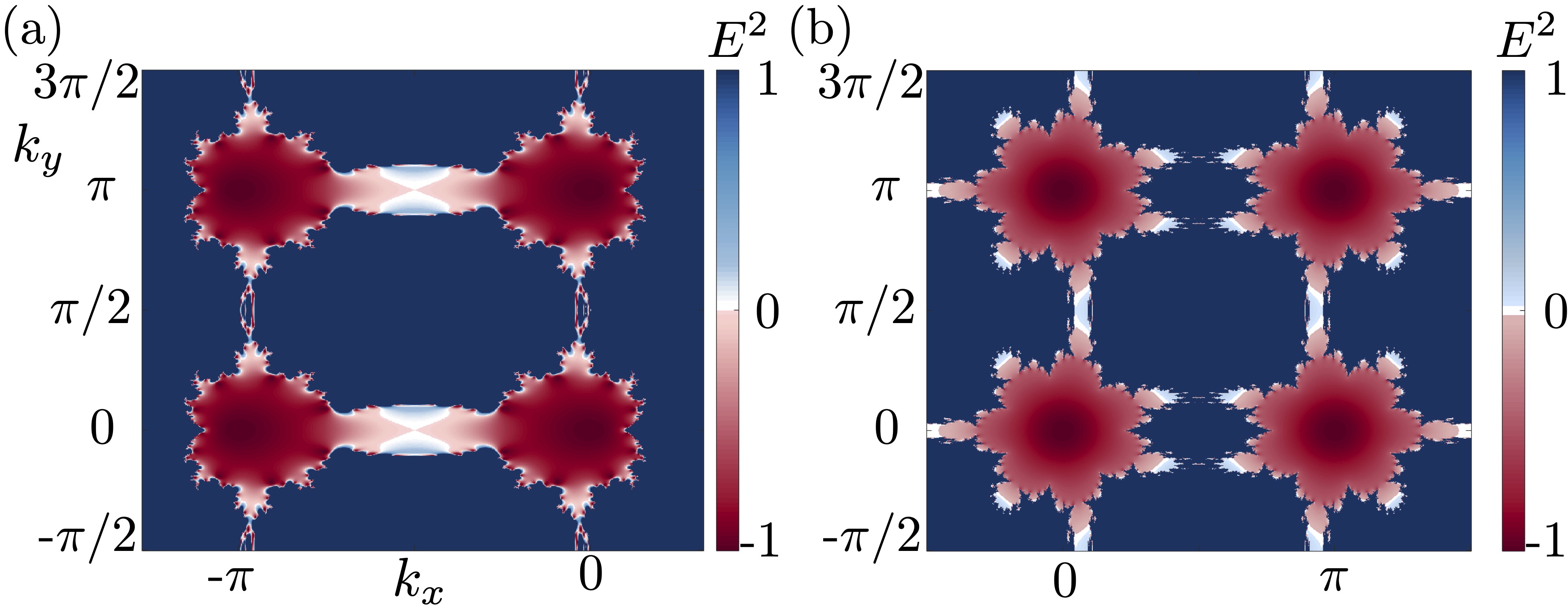}

\caption{\label{fig:PTper}Fermi surfaces [defined as $\text{Re}\left(E\right)=0 \iff E^2<0$] and their bounding EPs of the $\PT$-symmetric model given by Eqs.~\eqref{eq:drxPT2dper}-\eqref{eq:dryPT2dper} with (a) $d=2$ and (b) $d=8$, taking fractallike forms. In both panels, $r=2$ and Eq.~\eqref{eq:iteration} is terminated after ten steps.}
\end{figure}

Apart from the polynomial models presented in the main text and above, fractals are also of relevance in periodic models. One such example is obtained by thinking about each momentum component as an approximation of the corresponding sine-function, which would directly yield a periodic model. To illustrate this, we consider a $\mathcal{PT}$-symmetric two-band Hamiltonian on the form $\HH = d_{R,x}^p\sigma^x + d^p_{R,z}\sigma^z + i d^p_{I,y}\sigma^y$, with $d$-vector components given by
\begin{align}
    d_{R,x}^p &= \text{Re}\left[F(\sin k_x,\sin k_y,d)\right], \label{eq:drxPT2dper}
    \\
    d_{R,z}^p &= \text{Im}\left[F(\sin k_x,\sin k_y,d)\right],
    \\
    d_{I,y}^p &= \cos k_x+\cos k_y \label{eq:dryPT2dper}.
\end{align}
Here, we stress that similar results can be obtained for the other symmetries discussed above by identical reasoning. We recall that $d_{R,x}^p, d_{R,z}^p$ and $d_{I,y}^p$ are now real-valued periodic functions of the lattice momentum. The corresponding EP2s and FSs are displayed in Fig.~\ref{fig:PTper}. Despite being less prominent than their polynomial counterpart, the exceptional structures are undoubtedly fractallike also in these systems, making a more direct connection to experimental setups. It should be stressed, however, that the models presented in Fig.~\ref{fig:PTper} are indeed complicated to obtain in actual experiments, as the powers of the trigonometric functions in the Hamiltonian [which will be as high as 18 in the model presented in Fig.~\ref{fig:PTper}(a)] represent hopping distance on a lattice.

\end{document}